\documentclass[aps,pra,twocolumn,showpacs,superscriptaddress]{revtex4-1}

\usepackage{graphicx}
\usepackage[usenames]{color}
\usepackage{amssymb,amsmath}

\newcommand{\eq}[1]{Eq.~(\ref{#1})}
\newcommand{\fig}[1]{Fig.~\ref{#1}}
\newcommand{\be}[1]{\begin{equation}\label{#1}}
\newcommand{\ee}{\end{equation}}

\begin{document}

\title{
Auger spectra following inner-shell ionization of Argon by a Free-Electron Laser}

\author{A. O. G. Wallis}
\affiliation{Department of Physics and Astronomy, University College London, Gower Street, London WC1E 6BT, United Kingdom}
\author{L. Lodi}
\affiliation{Department of Physics and Astronomy, University College London, Gower Street, London WC1E 6BT, United Kingdom}
\author{A. Emmanouilidou}
\affiliation{Department of Physics and Astronomy, University College London, Gower Street, London WC1E 6BT, United Kingdom}

\date{\today}

\begin{abstract}
We explore the possibility of retrieving   Auger spectra with FEL radiation. Using a laser pulse of 260 eV photon energy, we study the interplay of photo-ionization and Auger processes following the initial formation of a 2p inner-shell hole in Ar.  Accounting for the fine structure of the ion states we demonstrate how to retrieve the Auger spectrum  of $\mathrm{Ar^{+}\rightarrow Ar^{2+}}$. Moreover, considering two electrons in coincidence we also demonstrate how to retrieve the Auger spectrum of $\mathrm{Ar^{2+}\rightarrow Ar^{3+}}$.

\end{abstract}


\maketitle

\section{Introduction}
The response of atoms to intense extreme ultraviolet (XUV) and X-ray  Free-Electron-Lasers  (FEL) is a fundamental theory problem. In addition, understanding FEL-driven processes  is of interest  for accurate modeling of laboratory and astrophysical plasmas. The fast progress in generating  intense FEL pulses of femtosecond duration renders timely the study of FEL driven processes in atoms. Such processes include the formation of inner-shell vacancies by photo-absorption and the subsequent Auger decays. Exploring the interplay of  photo-ionization and Auger processes is  a key to  understanding the rich electron dynamics underlying the formation 
  of highly charged ions \cite{Sorokin:2007,Young:Ne:2010,Doumy:nonlinear:2011} and hollow atoms  \cite{Young:Ne:2010,Fukuzawa:2013,Frasinski:2013}.

Auger spectra have attracted a lot of interest over the years with early studies  involving the formation of an  inner-shell hole following the impact of a particle, such as an electron \cite{mehlhorn1985atomic,Mcguire:nobelgas:1975,Mcguire:argon:1975,Mehlhorn:1968,Werme:1973}. 
From the early 80s, synchrotron radiation has largely replaced particle impact as a triggering mechanism of Auger processes  \cite{vonBusch:1994,Alkemper:1997,vonBusch:satellite:1999,Lablanquie:multi:2011}. 
Such studies include the detailed Auger spectrum following the decay of Ar$^+(2p^{-1})$  \cite{Pulkkinen:1996,Lablanquie:2007}.
The reason for using synchrotron radiation  is that it is monochromatic and allows for well defined initial excitations in the soft and hard X-ray regime. 
A recent study with synchrotron radiation \cite{Huttula:2013} involves the measurement of Auger spectra following the decay of the $\mathrm{Ar}^{2+}(2p^{-1}v^{-1})$  ionic states;  $v^{-1}$ is a hole in a valence orbital and $\mathrm{Ar}^{2+}(2p^{-1}v^{-1})$ is formed  by single-photon double ionization. 

In this work, we explore the feasibility of obtaining detailed Auger spectra using FEL radiation. FEL radiation  allows for well-defined initial excitations. It also allows for the creation of multiple inner-shell holes resulting in  multiple  Auger decays; generally the Auger spectra thus generated have larger yields than those generated from synchrotron radiation. The increasing availability of FEL sources provides an additional motivation for the current study.
We explore the interplay of photo-ionization and Auger processes in $\mathrm{Ar}$ interacting with a 260 eV FEL pulse, a photon energy  sufficient to ionize a single inner-shell $\mathrm{2p}$ electron in Ar. We 
compute the ion yields due to Auger and photo-ionization processes  and study the ion yields dependence  on the FEL pulse parameters.     To do so we solve a set of rate equations \cite{Rohringer:2007,Makris:2009}. Initially, in the rate equations we only  account for the electronic configuration of the ion states. This simplification allows us to gain insight into the processes involved and explore the optimal parameters for observing Auger spectra. We next proceed to fully account for the fine structure of the ion states in the rate equations. We subsequently obtain the detailed  Auger spectrum of $\mathrm{Ar^{+}\rightarrow Ar^{2+}}$. Moreover,  we  demonstrate how the detailed  Auger spectrum of $\mathrm{Ar^{2+}\rightarrow Ar^{3+}}$   can be observed in an FEL two-electron coincidence experiment.

\section{Auger and Ion yields excluding fine structure }

We model the response of Ar to a 260 eV  FEL pulse by formulating and solving a set of rate equations for the time  dependent populations of the ion states \cite{Rohringer:2007,Makris:2009}.  Our first goal is to gain insight into how the ion and Auger yields depend on the duration and intensity of the laser pulse. To do so, in this section, we simplify the theoretical treatment by  accounting only for the electronic configuration, i.e, $(1s^a, 2s^b, 2p^c, 3s^d, 3p^e)$ of the ion states and not the fine structure of these states. 
%
%
By fine structure we refer to all possible $^{2S+1}L_J$ states for a given electronic configuration, accounting for spin-orbit coupling.  
%
To compute the Auger transition rates between different electron configurations we use the formalism introduced by Bhalla {\it et al.} \cite{Bhalla:1973} and refer to these transition rates as Auger group rates in accord with \cite{Bhalla:1973}.

\begin{figure}
\centerline{\includegraphics[width=0.99\linewidth]{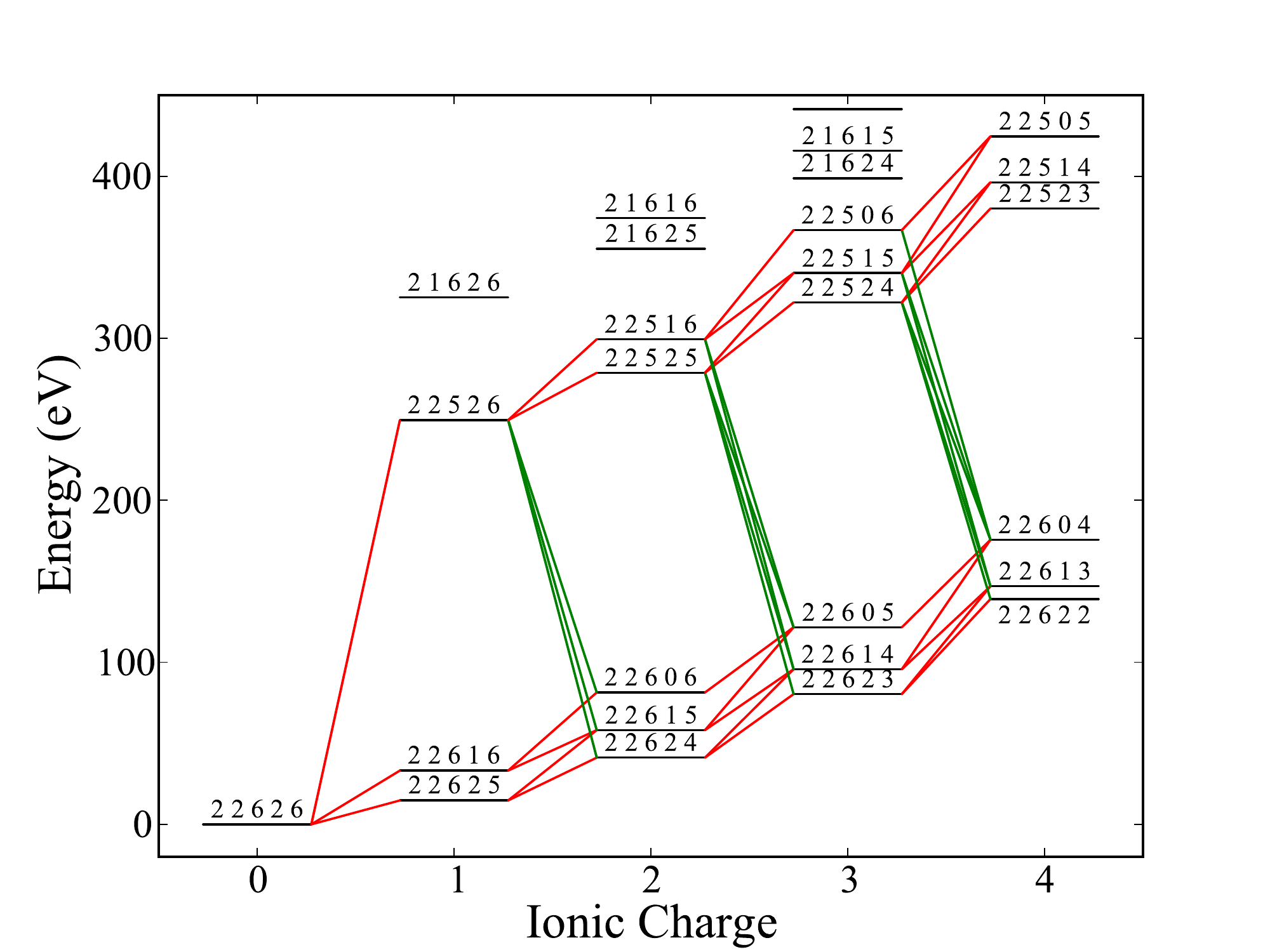}}
\caption{\label{fig:Ar_levels}(Color online) Ionization pathways between different electronic configurations of Ar, up to Ar$^{4+}$, accessible with sequential single-photon ($\mathrm{\hbar\omega=260 eV})$ absorptions and Auger decays. 
The label  $\mathrm{(a,b,c,d,e)}$ stands for the electronic configuration  $\mathrm{(1s^a, 2s^b, 2p^c, 3s^d, 3p^e)}$. The red and green lines indicate photo-ionization and Auger transitions, respectively.}
\end{figure}

\subsection{Rate equations}

In the rate equations 
we account for single-photon ionization  and Auger transitions. For the ion states considered  the X-ray fluorescence widths are typically three orders of magnitude smaller than the Auger decay widths \cite{Chen:1974}; we can thus safely neglect the former. In \fig{fig:Ar_levels}, accounting for states up  to Ar$^{4+}$, we illustrate the photo-ionization and Auger  transitions between states with different electron configurations that are allowed for a laser pulse of 260 eV photon energy. This photon energy  is sufficient for creating a single inner-shell $\mathrm{2p}$ hole and multiple valence holes in Ar. In the rate equations we  include all possible ion states accessible by a 260 eV laser-pulse; the highest ion state is
$\mathrm{Ar^{9+}(1s^2,2s^2,2p^5,3s^0,3p^0)}$. 
The rate equations describing the population $\mathcal{I}_i^{(q)}$ of an ion state $\mathrm{i}$ with charge $\mathrm{q}$ take the form

\begin{align}\label{eqn:rate1}
\frac{d}{dt}&\mathcal{I}_j^{(q)}(t) = 
 \sum_i  \left(\sigma_{i\rightarrow j} J(t)+ \Gamma_{i\rightarrow j} \right) \mathcal{I}_i^{(q-1)}(t) 
  \\ \nonumber
 & -  \sum_k \left( \sigma_{j\rightarrow k} J(t) + \Gamma_{j\rightarrow k} \right)  \mathcal{I}_j^{(q)}(t) 
 \\ \nonumber
 \frac{d}{dt} &\mathcal{A}^{(q)}_{i\rightarrow j} = \Gamma_{i\rightarrow j} \mathcal{I}_i^{(q-1)}(t),\\ \nonumber
 \frac{d}{dt}&\mathcal{P}^{(q)}_{i\rightarrow j}= \sigma_{i\rightarrow j} J(t) \mathcal{I}_i^{(q-1)}(t), 
\end{align}
where $\mathrm{\sigma_{i\rightarrow j}}$ and $\mathrm{\Gamma_{i\rightarrow j}}$  are the single-photon absorption cross section and Auger decay rate from initial state $\mathrm{i}$ to final state $\mathrm{j}$, respectively. $\mathrm{J(t)}$ is the photon flux. Atomic units are used in this work.
The temporal form of the FEL flux is modelled with a Gaussian function \cite{Rohringer:2007} which is given by
\begin{equation}
J(t) = 1.554\times 10^{-16} \frac{I_0\textrm{[W cm}^{-2}]}{\hbar\omega\textrm{[eV]}} \exp\left\{ -4\ln2 \left(\frac{t}{\tau_X}\right)^2 \right\}
\end{equation}
with $\mathrm{\tau_X}$ the full width at half maximum  and $\mathrm{I_0}$ the peak intensity.
The first term in \eq{eqn:rate1} accounts for the formation of the state $\mathrm{j}$ with charge $\mathrm{q}$ through the single-photon ionization and Auger decay of the state $\mathrm{i}$ with charge $\mathrm{q-1}$. The second term in \eq{eqn:rate1}  
 accounts for the depletion of state $\mathrm{j}$ by single-photon ionization and Auger decay to the state $\mathrm{k}$ with charge $\mathrm{q+1}$.
In \eq{eqn:rate1}, we also solve for the Auger yield $\mathcal{A}^{(q)}_{i\rightarrow j}$ from an initial state $\mathrm{i}$ with charge $\mathrm{q-1}$ to a final state $\mathrm{j}$ with charge $\mathrm{q}$. In addition, we solve for  the photo-ionization yield $\mathcal{P}^{(q)}_{i\rightarrow j}$
from an initial state $\mathrm{i}$ with charge $\mathrm{q-1}$ to a final state $\mathrm{j}$ with charge $\mathrm{q}$. These yields provide the probability for observing an electron with energy corresponding to the transition $\mathrm{i\rightarrow j}$. The total Auger and photo-ionization yields for the transition from any state with charge $\mathrm{q-1}$ to any state with charge $\mathrm{q}$ are given by
\begin{align}
\mathcal{A}^{(q)} = \sum_{i,j} \mathcal{A}^{(q)}_{i\rightarrow j}, & &
\mathcal{P}^{(q)} = \sum_{i,j} \mathcal{P}^{(q)}_{i\rightarrow j} .
\end{align}
To find the total ion yield of a state with charge $\mathrm{q}$, i.e., the ion yield for   Ar$^{q+}$ we  sum over the populations of all ion states  with charge $\mathrm{q}$
\begin{equation}
\mathcal{I}^{(q)} = \sum_i \mathcal{I}_i^{(q)}.
\end{equation}
All yields are computed long after the end of the pulse.

As we show later in the paper, it is also of interest to compute the Auger and photo-ionization yields  along a pathway $\mathrm{i \rightarrow j \rightarrow k}$. These yields provide the probability for observing in a two-electron coincidence experiment  two electrons with energies corresponding to the transitions $\mathrm{i\rightarrow j}$ and  $\mathrm{j\rightarrow k}$. If there is only one state $\mathrm{i}$  leading to state $\mathrm{j}$ then the probability for observing the electron emitted in the transition $\mathrm{i\rightarrow j}$ and the electron emitted in the transition $\mathrm{j\rightarrow k}$ is simply the Auger $\mathcal{A}^{(q)}_{j\rightarrow k}$ or the photo-ionization $\mathcal{P}^{(q)}_{j\rightarrow k}$ yield. However, it can be the case that  we have multiple states leading to state $\mathrm{j}$, for example, $\mathrm{i \rightarrow j \rightarrow k}$ and $\mathrm{i' \rightarrow j \rightarrow k}$. Then to compute the probability $\mathcal{P}^{(q)}_{j(i) \rightarrow k}$ or $\mathcal{A}^{(q)}_{j(i) \rightarrow k}$   for observing the electron emitted in the transition $\mathrm{i\rightarrow j}$ and the electron emitted in the transition $\mathrm{j\rightarrow k}$ we need to solve separately for the contribution of state $\mathrm{i}$ to the population of state $\mathrm{j}$:
\begin{align}\label{eqn:coincidence}
\frac{d}{dt} \mathcal{I}^{(q-1)}_{j(i)}(t) =
& (\sigma_{i\rightarrow j}J(t)+\Gamma_{i\rightarrow j})\mathcal{I}^{(q-2)}_{i}(t) \\ \nonumber 
&- \sum_{k'}(\sigma_{j\rightarrow k'}J(t)+\Gamma_{j\rightarrow k'})\mathcal{I}^{(q-1)}_{j(i)}(t) \\ \nonumber
\frac{d}{dt} \mathcal{P}^{(q)}_{j(i) \rightarrow k} = & \sigma_{j\rightarrow k}J(t)\mathcal{I}^{(q-1)}_{j(i)}(t) \\ \nonumber
\frac{d}{dt} \mathcal{A}^{(q)}_{j(i) \rightarrow k} = & \Gamma_{j\rightarrow k}\mathcal{I}^{(q-1)}_{j(i)}(t). \end{align}

%
%
\subsection{Auger group rates}

To compute the Auger group rates $\Gamma_{i\rightarrow j}$ we use the formulation of Bhalla et al. \cite{Bhalla:1973}. For each electron configuration included in the rate equations, we obtain the energy and bound atomic orbital  with a Hartree-Fock (HF) calculation. These calculations are performed with the \textit{ab initio} quantum chemistry package \textsc{molpro} \cite{MOLPRO_brief_2009.1} using the split-valence 6-311G basis set. 
To compute the continuum orbital that describes the outgoing Auger electron we use the Hartree-Fock-Slater (HFS) 
one-electron potential that is obtained using an updated version of the Herman Skillman atomic structure code \cite{HermanSkillman:1963,hermsk:program}. 
This one electron potential is expressed in terms of an effective nuclear charge $\mathrm{Z_\text{HFS}(r)}$. The resulting radial HFS  equation is of the form
\begin{equation}\label{eqn:HFSwavefnc}
\left[ -\frac{d^2}{dr^2} +\frac{l(l+1)}{r^2} -\frac{Z_\text{HFS}(r)}{r} \right]P_{nl}(r) = E P_{nl}(r),
\end{equation}
where the orbital wavefunction is given by $\mathrm{\psi_{nlm}(\mathbf{r})= r^{-1}P_{nl}(r)Y_{lm}(\hat{r})}$. We solve equation \eq{eqn:HFSwavefnc} for the continuum orbital ($\mathrm{E>0}$)
using the modified Numerov method \cite{Numerov:1933,Melkanoff:1966}. We match the solution to the appropriate asymptotic boundary conditions for energy normalized continuum wave functions \cite{Child:1974}.  
In Table I we list our results for the Auger group rates $\mathrm{Ar^{+}(2p^{-1})\rightarrow Ar^{2+}(3s^{-1}3p^{-1})}$, $\mathrm{Ar^{+}(2p^{-1})\rightarrow Ar^{2+}(3s^{-2})}$ and $\mathrm{Ar^{+}(2p^{-1})\rightarrow}$  $\mathrm{Ar^{2+}(3p^{-2})}$ and compare them with two other calculations that employ the HFS method  \cite{McGuire:1971} and  the  HF method  \cite{Dyall:1982} both for the bound and the continuum orbitals. As expected, 
our results lie between the results of these two calculations. For reference, we also list in Table I  the results from a  Configurational Interaction (CI) calculation  \cite{Dyall:1982}. In Table II we list our results for all the Auger group rates involved in the rate equations for Ar for a 260 eV FEL pulse.

\begin{table}
\caption{Auger group rates  for a transition from an initial state  $\mathrm{(1s^a,2s^b,2p^c,3s^c,3p^e)}$ 
to a final state where  the electron filling in the $\mathrm{2p}$ hole in the initial state and the electron escaping to the continuum  occupy $\mathrm{nl}$ and $\mathrm{n'l'}$ orbitals. We also list the Auger rates 
 obtained in \cite{McGuire:1971} using the Hartree-Fock-Slater (HFS) method,  in \cite{Dyall:1982} using a Hartree-Fock (HF) method, and in \cite{Dyall:1982} using a CI calculation. The rates are given in 10$^{-4}$ a.u. }
\begin{center}
\begin{tabular}{p{2.5mm} p{2.5mm} p{2.5mm} p{2.5mm} p{2.5mm} | c | c c   c c}
\hline
\multicolumn{5}{c|}{initial config.} & method &  \multicolumn{4}{c}{group rates ($10^{-4}$ a.u.)} \\
a & b & c & d & e & &$3s3s$ & $3s3p$ & $3p3p$ & total \\
\hline \hline
2 & 2 & 5 & 2 & 6 & HFS \cite{McGuire:1971} & 0.77 & 12.85 & 47.90 & 61.52 \\
   &    &    &   &     & HF \cite{Dyall:1982} & 0.28 & 15.74 & 56.97 & 72.99 \\
   &    &    &   &     & CI  \cite{Dyall:1982} & 0.47 &  9.54 & 54.74 & 64.75 \\
   &    &    &   &     & this work & 0.45 & 15.60 & 51.67 &  67.72 \\
\hline
\end{tabular}
\end{center}
\end{table}

\begin{table}
\caption{As in Table I for results obtained in this work for all Auger group rates included in the rate equations.
\label{tab:HFgroupRates}
}
\begin{center}
\begin{tabular}{p{2.5mm} p{2.5mm} p{2.5mm} p{2.5mm} p{2.5mm} |  c c  c c}
\hline
\multicolumn{5}{c|}{initial config.} & \multicolumn{4}{c}{group rates ($10^{-4}$ a.u.)} \\
a & b & c & d & e & $3s3s$ & $3s3p$ & $3p3p$ & total   \\ \hline \hline
2&2&5&2&6&0.450&15.598&51.665&67.713 \\
\hline
2&2&5&2&5&0.502&9.615&25.457&35.575 \\
2&2&5&1&6&-&9.244&58.693&67.937   \\
\hline
2&2&5&2&4&0.568&9.429&20.324&30.321 \\
2&2&5&1&5&-&5.780&29.273&35.053 \\
2&2&5&0&6&-&-&68.708&68.708 \\
\hline
2&2&5&2&3&0.638&7.973&11.680&20.291   \\
2&2&5&1&4&-&5.631&23.952&29.583 \\
2&2&5&0&5&-&-&33.761&33.761  \\
\hline
2&2&5&2&2&0.710&5.845&4.349&10.905  \\
2&2&5&1&3&-&4.650&13.337&17.986\\
2&2&5&0&4&-&-&23.946&23.946 \\
\hline
2&2&5&2&1&0.778&2.843&-&3.621  \\
2&2&5&1&2&-&3.374&4.909&8.283  \\
2&2&5&0&3&-&-&14.309&14.309 \\
\hline
2&2&5&2&0&0.863&-&-&0.863  \\
2&2&5&1&1&-&1.612&-&1.612 \\
2&2&5&0&2&-&-&5.168&5.168 \\
\hline
\end{tabular}
\end{center}
\end{table}

\subsection{Results for Auger and Ion yields}

\begin{figure}[h]
\begin{center}
\includegraphics[width=0.9\linewidth]{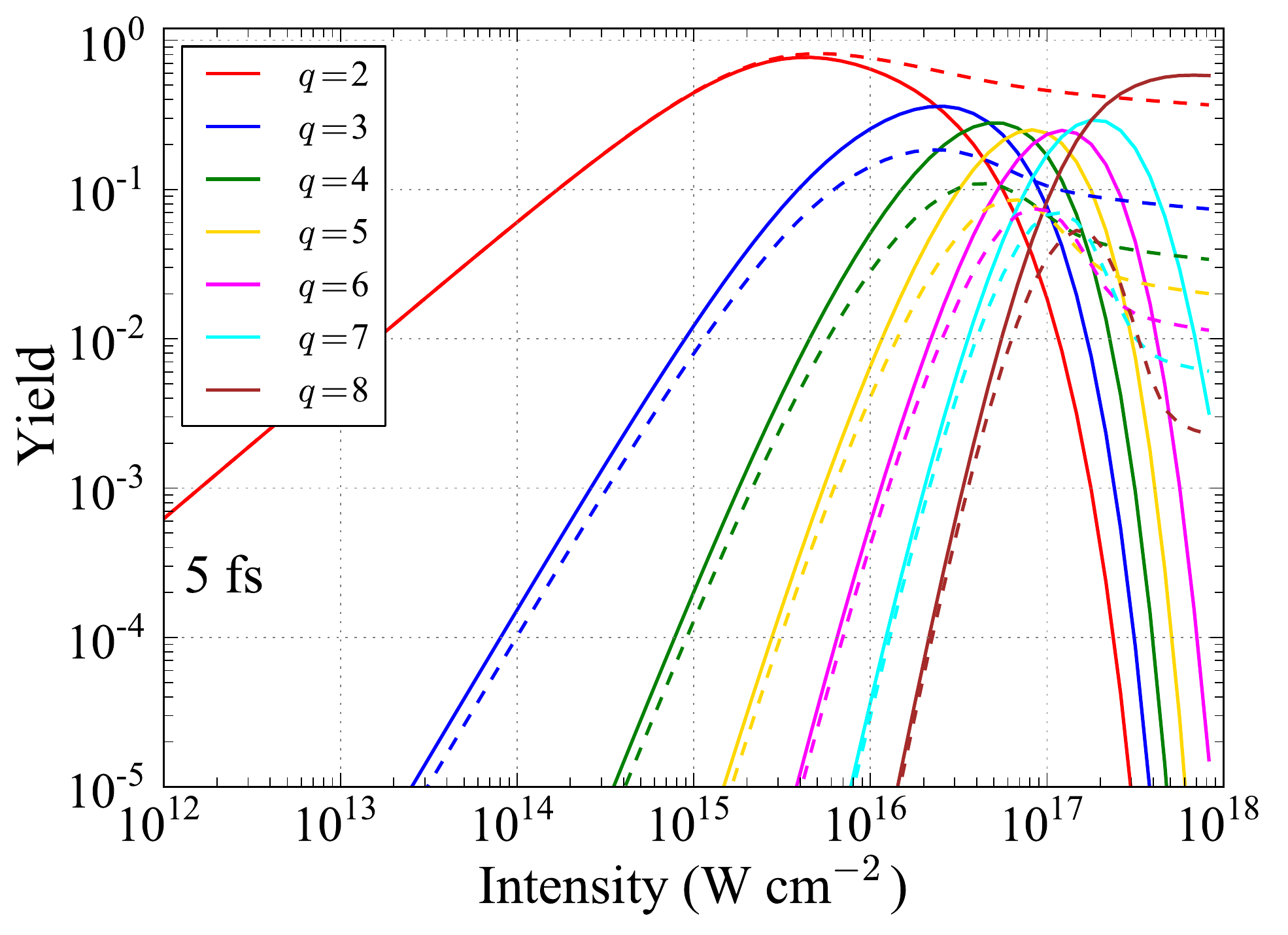}
\includegraphics[width=0.9\linewidth]{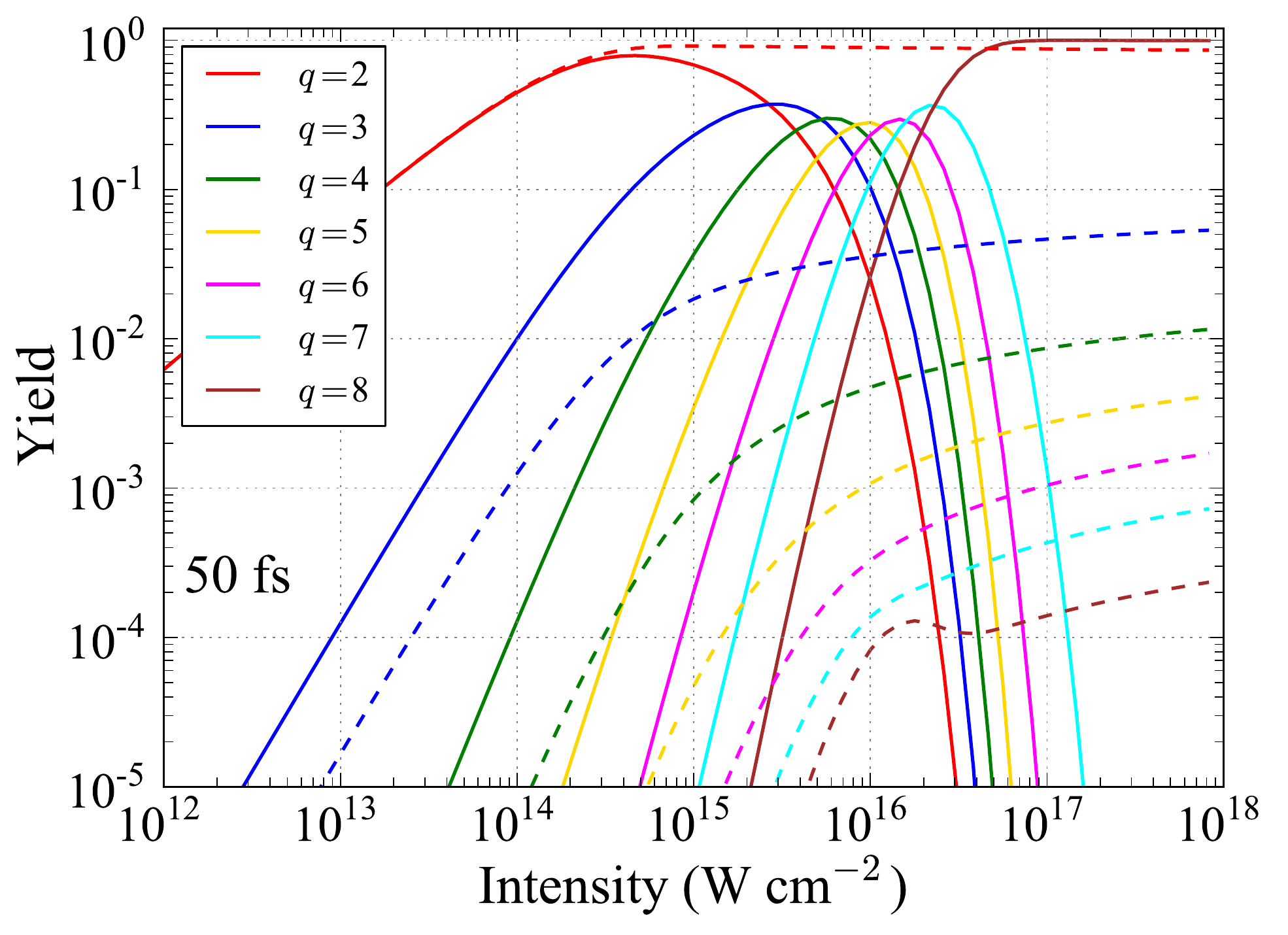}
\end{center}
\caption{(Color online) Total ion $\mathcal{I}^{(q)}$ (solid lines) and Auger $\mathcal{A}^{(q)}$  (dashed lines) yields as a function of intensity for pulse duration of 5 fs (top) and 50 fs (bottom).
\label{fig:Ar_GroupAugerIonYields}
}
\end{figure}

For the photo-ionization cross sections we use the Los Alamos National Laboratory atomic physics codes \cite{LANL:APC} that are based on the HF routines of R. D. Cowan \cite{Cowan:1981}. Assuming that the initial state is the neutral Ar, we solved numerically \cite{NumRec:2007} the set of first order differential rate equations in \eq{eqn:rate1}. In  \fig{fig:Ar_GroupAugerIonYields} we show our results for the total ion  $\mathcal{I}^{(q)}$ and Auger  $\mathcal{A}^{(q)}$ yields as a function of the pulse intensity for  pulse durations  of 5 fs and 50 fs. 
%
%
 From \fig{fig:Ar_GroupAugerIonYields} we observe that  $\mathcal{A}^{(q)}$ can be very similar to  $\mathcal{I}^{(q)}$ for  $q \ge 2$ depending on the pulse intensity and duration. Indeed, the formation of  $\mathrm{Ar^{q+}}$ occurs from a sequence of transitions where the final step involves either the single-photon ionization or the Auger decay of  $\mathrm{Ar^{(q-1)+}}$. For high pulse intensities, independent of the pulse duration, both final steps   are likely and thus $\mathcal{A}^{(q)}$ is different than  $\mathcal{I}^{(q)}$. For small pulse intensities, 
 if the pulse is short then the formation of $\mathrm{Ar^{q+}}$ through the Auger decay of $\mathrm{Ar^{(q-1)+}}$ is favored; if the pulse is long multi-photon absorption is highly likely  making possible formation of $\mathrm{Ar^{q+}}$ also through single-photon ionization of $\mathrm{Ar^{(q-1)+}}$. Thus, generally, for small pulse intensities, if the pulse is short $\mathcal{A}^{(q)} \approx \mathcal{I}^{(q)}$ while if the pulse is long $\mathcal{A}^{(q)} \ne \mathcal{I}^{(q)}$.   

\subsection{Truncation of the number of states included in the rate equations}

\begin{figure}[h]
\includegraphics[width=0.9\linewidth]{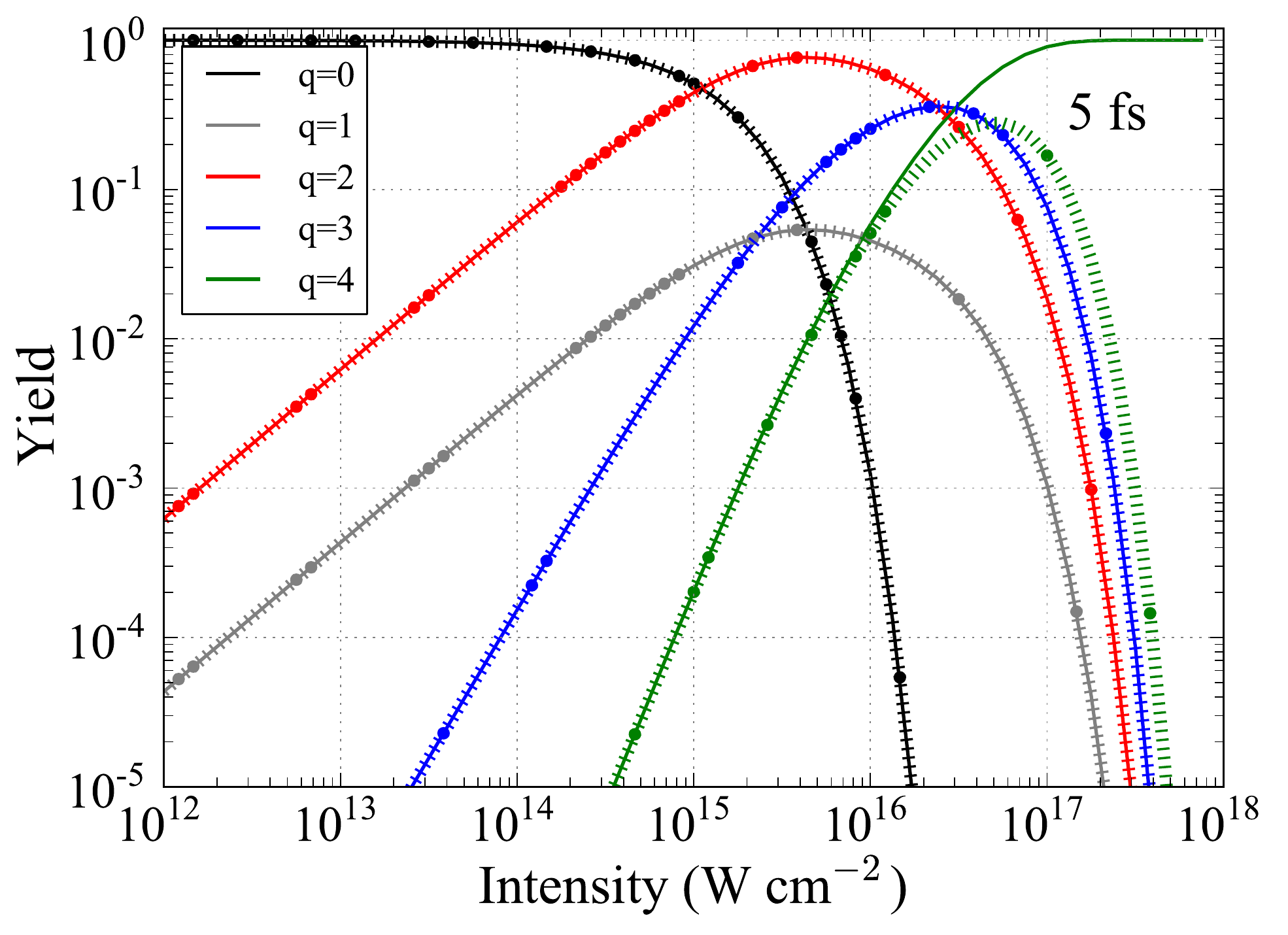}
\includegraphics[width=0.9\linewidth]{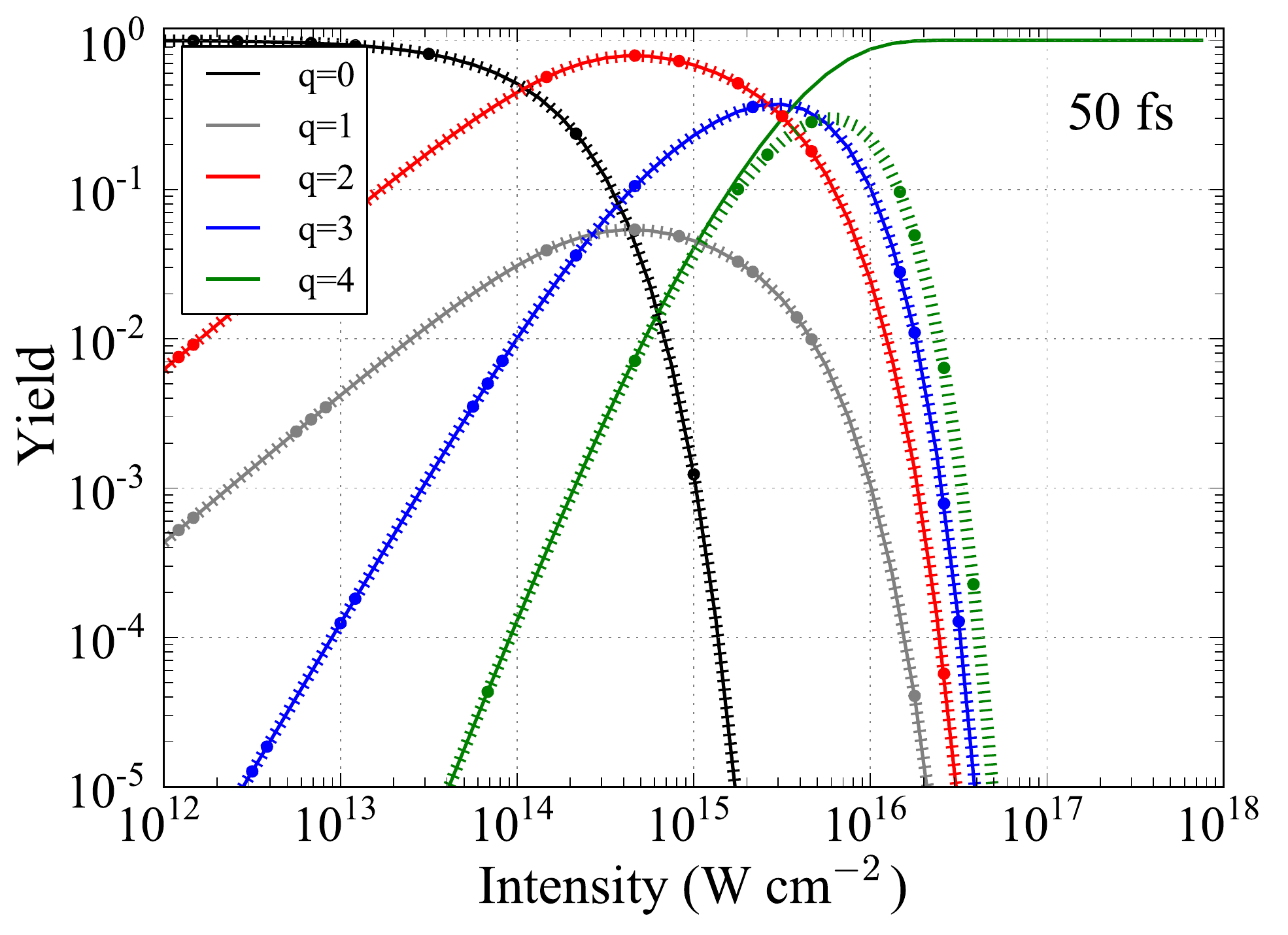}
\caption{(Color online) Total ion yields $\mathcal{I}^{(q)}$ for $q =$ 0,1,2,3,4 when  ion states up to Ar$^{9+}$ (dashed lines) and ion states up  to Ar$^{4+}$ (solid lines) are included as a function of pulse intensity for pulse durations 5 fs (top) and 50 fs (bottom).
\label{fig:truncation}}
\end{figure}

 \fig{fig:Ar_GroupAugerIonYields} shows that appropriate tuning of the laser parameters can result in large Auger yields even for high ion states. Regarding Auger spectra this is an advantage of FEL radiation compared to synchrotron radiation.  However, discerning the  Auger spectra produced by the FEL pulse  is a challenging task since many photo-ionization and Auger electrons escape to the continuum. In the next section
we focus on the Auger electron spectra resulting from ion states up to  Ar$^{3+}$. To accurately describe these spectra we need to account for the fine structure of the ion states included in the rate equations. However, such an inclusion results in a very  large increase  
of the number  of ion states  that need to be accounted for in the rate equations. For instance, when considering states up to Ar$^{4+}$ the number of ions states  in the rate equations  increases from  21(no fine structure)  to 186 (with fine structure). 
We thus truncate the number of ion states we consider.   In \fig{fig:truncation} we compare $\mathcal{I}^{(q)}$, for $\mathrm{q =}$ 1,2,3,4,  when we include  ion states up to Ar$^{9+}$ and up to Ar$^{4+}$. We find that the truncation affects only $\mathcal{I}^{(4)}$ while $\mathcal{I}^{(1)}$, $\mathcal{I}^{(2)}$ and $\mathcal{I}^{(3)}$ are unaffected. 
%
%
Since the focus of the current work is the Auger electron spectra up to Ar$^{3+}$, in what follows  we truncate to include only ion states  up to Ar$^{4+}$.
%
Moreover, comparing \fig{fig:truncation} with \fig{fig:Ar_GroupAugerIonYields}, we find that a pulse duration of 5 fs is  short enough for $\mathcal{A}^{(q)} \approx \mathcal{I}^{(q)}$  to be true for intensities up to roughly $10^{16}$ W cm$^{-2}$.  This guarantees that less photo-ionization electrons are ejected to the continuum making it easier to discern the Auger electrons. We also find that for pulse intensities around $10^{15}$-$10^{16}$ W cm$^{-2}$ both $\mathcal{A}^{(2)}$ and $\mathcal{A}^{(3)}$ yields have significant values. Thus, a laser pulse with duration of 5 fs  and intensity of $5\times10^{15}$ W cm$^{-2}$ is optimal for the experimental observation of the Auger electron spectra up to Ar$^{3+}$.

\section{Auger spectra }
 
 \subsection{Computation of fine structure ion states}
 
We next describe the method we use  to compute the fine  structure states of  each electron configuration that is included in the truncated rate  equations. 
To obtain the fine structure ion states we use the \textsc{grasp2k} package \cite{grasp2k:2013} and the \textsc{relci} extension \cite{RELCI:2002} provided in the \textsc{ratip} package \cite{RATIP:2012}. These packages are used to perform relativistic calculations within the Multi-Configuration Dirac-Hartree-Fock (MCDHF) formalism \cite{Grant:book:2006}. The photo-ionization cross sections and Auger decay rates between fine structure states are then calculated  using the \textsc{photo} and \textsc{auger} components of the \textsc{ratip} package. 
Since \textsc{grasp2k} utilizes the Dirac equation the calculations are performed in the $j$-$j$ coupling scheme.
We briefly outline the steps we follow  to obtain the fine structure states for a given electron configuration of Ar; where appropriate we  illustrate using  Ar$^+(1s^2,2s^2,2p^5,3s^2,3p^6)$.

1) We identify the fine structure states for the electron configuration at hand; in our example these states are $\mathrm{^{2}P_{1/2}}$ and $\mathrm{^{2}P_{3/2}}$. We identify the configurational state functions (CSFs) that can be  constructed out of the possible $\mathrm{nlj}$ orbitals; in our example the possible CSFs are
$$1. (1s_{1/2}^2,2s_{1/2}^2,2p_{1/2}^{1},2p_{3/2}^4,3s_{1/2}^2,3p_{1/2}^2,3p_{3/2}^4); J^P=\frac{1}{2}^{-}$$
$$2. (1s_{1/2}^2,2s_{1/2}^2,2p_{1/2}^{2},2p_{3/2}^3,3s_{1/2}^2,3p_{1/2}^2,3p_{3/2}^4); J^P=\frac{3}{2}^{-}$$
Each fine structure state is  a linear combination of the CSFs that have the same total angular momentum $J$ and parity $P$; in our example $\mathrm{^2P_{1/2}}$ is expressed in terms of the first CSF and  $\mathrm{^{2}P_{3/2}}$ in terms of the second CSF. A Self-Consistent-Field (SCF) DHF calculation is now performed for all the CSFs. This calculation optimizes the $\mathrm{nlj}$ orbitals and the coefficients in the expansion of each  fine structure state in terms of CSFs. 

2) To account for electron correlation, as a first step, we include the additional orbitals $\mathrm{3d_{3/2}}$ and $\mathrm{3d_{5/2}}$. A new set of CSFs is generated from the single and double excitations of the step-1 CSFs, while keeping the occupation of the $\mathrm{1s}$, $\mathrm{2s}$ and $\mathrm{2p}$ orbitals frozen. A new MCDHF calculation is then performed with the new set of CSFs keeping the step-1 $\mathrm{nlj}$ orbitals frozen and only optimizing the newly added ones.

3)  As a second step in accounting for electron correlation, we include all orbitals up to $\mathrm{4d_{3/2}}$, $\mathrm{4d_{5/2}}$. Again, as for step-2, a new set of CSFs is generated from the single and double excitations of the step-1 CSFs, while keeping the occupation of the $1s$, $2s$ and $2p$ orbitals frozen. Another MCDHF calculation is performed optimizing only the newly added, compared to step-2, orbitals. 
Introducing correlation orbitals layer by layer as described in steps 1-3 is the recommended procedure  in the \textsc{grasp2k} manual in order to achieve convergence of the SCF calculations.

4) Finally, using the orbitals generated in steps 1-3 we perform a CI calculation that optimizes the coefficients that express each fine structure state in terms of all the CSFs generated in steps 1-3.

\begin{table}
\caption{The Auger  energies and rates from the $\mathrm{Ar(2p^{-1}_{1/2}})$ and $\mathrm{Ar(2p^{-1}_{3/2}})$  initial state to a final state where  the electron filling in the $\mathrm{2p}$ hole and the electron escaping to the continuum  occupy $\mathrm{nl}$ and $\mathrm{n'l'}$ orbitals.
We list the MCDHF results obtained in this work and  the experimental results of Pulkkinen {\it et al}. \cite{Pulkkinen:1996}.
The Auger electron energies $\mathrm{E}$ are given in eV and the Auger rates $\Gamma$ are given in $10^{-4}$ a.u.
\label{tab:Ar2p_CI}
}
\begin{center}
\begin{tabular}{c c | c c | cc c}
\hline
\multicolumn{2}{c|}{Final State} &\multicolumn{2}{c|}{Exp. \cite{Pulkkinen:1996} }&\multicolumn{3}{c}{this work } \\
 & & $E$ & $I$ & $E$ & $\Gamma$ & $I$  \\
\hline \hline
\multicolumn{7}{c}{ Ar$^+(2p_{1/2}^{-1}$) }  \\
\hline
$3p3p$
   	&$^3P_2$&207.39&76&207.57&2.37&64 \\
 &$^3P_1$&207.25&176&207.44&5.11&138 \\
 &$^3P_0$&207.20&60&207.38&2.13&58 \\
 &$^1D_2$&205.65&404&205.64&11.78&318 \\
 &$^1S_0$&203.26&100&203.35&3.70&100 \\
$3s3p$
	&$^3P_2$&-&-&193.25&0.02&1 \\
 &$^3P_1$&193.13&24&193.12&1.12&30 \\
 &$^3P_0$&193.07&18&193.06&0.58&16 \\
 &$^1P_1$&189.50&39&188.66&1.85&50 \\
$3s3s$
	&$^1S_0$&176.43&6&175.36&0.62&17 \\
\hline 
\multicolumn{7}{c}{Ar$^+(2p_{3/2}^{-1}$) } \\
\hline
$3p3p$
	&$^3P_2$&205.24&261&205.43&7.58&240 \\
 &$^3P_1$&205.10&73&205.30&2.77&88 \\
 &$^3P_0$&205.08&26&205.24&0.73&23 \\
 &$^1D_2$&203.50&390&203.50&11.01&348 \\
 &$^1S_0$&201.11&100&201.22&3.16&100 \\
$3s3p$
	&$^3P_2$&191.09&77&191.11&1.52&48 \\
 &$^3P_1$&190.95&11&190.98&0.35&11\\
 &$^3P_0$&-&-&190.92&0&0 \\
&$^1P_1$&187.39&71&186.52&1.79&57 \\
$3s3s$
	&$^1S_0$&174.27&13&173.22&0.61&19 \\
\hline
\end{tabular}
\end{center}
\end{table}%

In Table ~\ref{tab:Ar2p_CI} we list the energies and Auger rates we obtain using the method described above for the fine structure states of Ar$^{+}(2p^{-1})$. To directly compare with the experimental results in \cite{Pulkkinen:1996} we define the intensity for an Auger decay from an initial state $\mathrm{i}$ to a final state $\mathrm{j}$ as 
\begin{equation}
I_{i\rightarrow j} = \frac{ \Gamma_{i \rightarrow j} }{ \sum_j  \Gamma_{i \rightarrow j} },
\end{equation}
and is scaled such that the intensity for the transition $\mathrm{Ar^+(2p^{-1})\rightarrow Ar^{2+}(3p^{-2}; ^1S_0)}$ is equal to 100 in accord with \cite{Pulkkinen:1996}. It can be seen that our calculated results are in good agreement with the experimental results of Pulkkinen {\it et al}. \cite{Pulkkinen:1996}.

\subsection{Results for Auger and Ion yields including fine structure}

\begin{figure}[h]
\begin{center}
\includegraphics[width=0.9\linewidth]{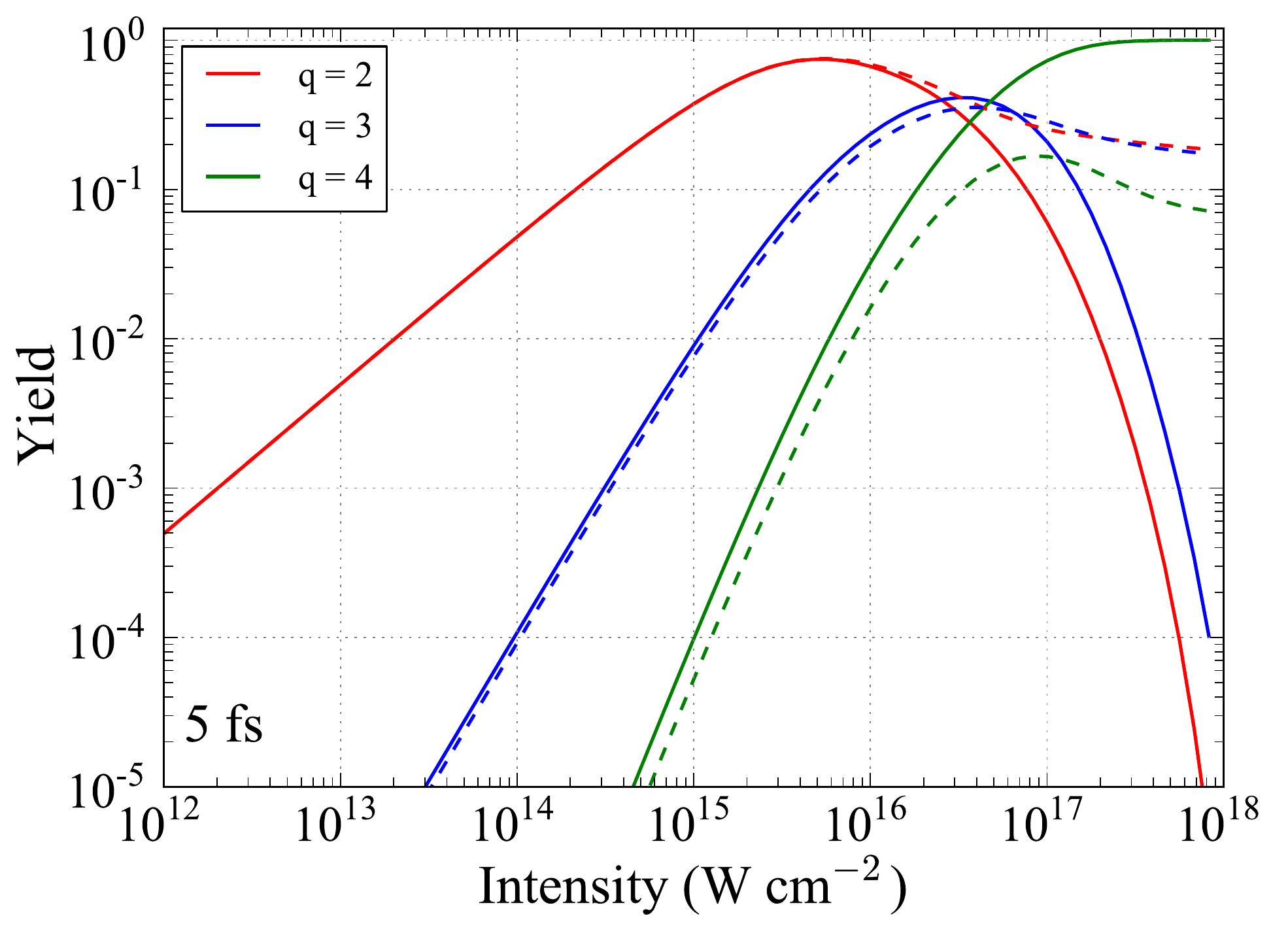}
\end{center}
\caption{(Color online) Total ion $\mathcal{I}^{(q)}$ (solid lines) and Auger $\mathcal{A}^{(q)}$ yields (dashed lines) for $q=2,3,4$ as a function of intensity for a pulse duration of 5 fs.
These yields are calculated  with fine structure included in the rate equations.
\label{fig:FS_AugerIon}}
\end{figure}

\begin{figure}[h]
\centerline{\includegraphics[width=0.9\linewidth]{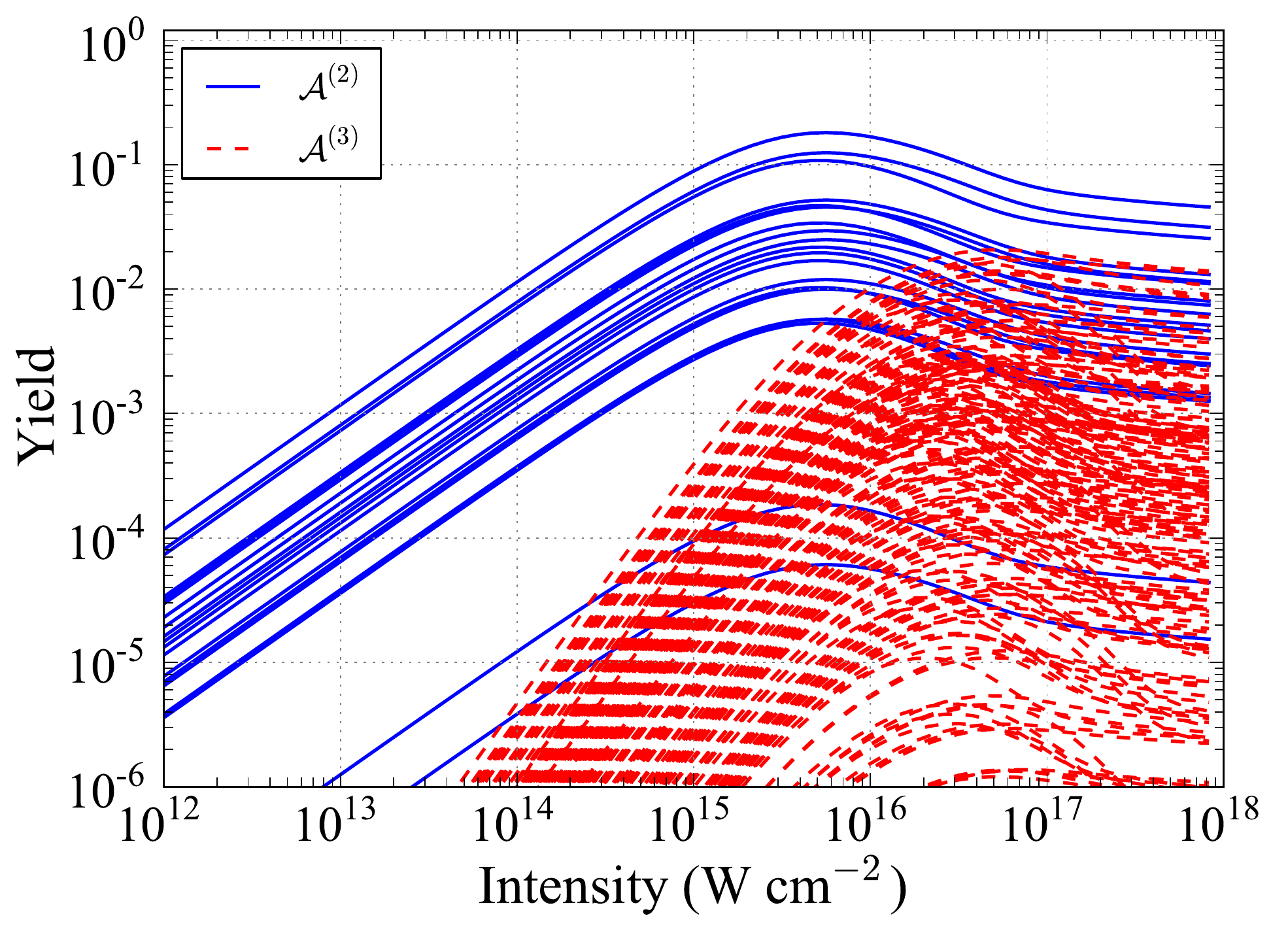}}
\caption{\label{fig:FS_Auger}
(Color online) The Auger yields $\mathcal{A}^{(2)}_{i\rightarrow j}$ (blue, solid lines)  and $\mathcal{A}^{(3)}_{i\rightarrow j}$ (red, dashed lines)  as a function of intensity for a pulse duration of 5 fs.
These yields are calculated with fine structure included in the rate equations.}
\end{figure}

In \fig{fig:FS_AugerIon} we show the total ion  $\mathcal{I}^{(q)}$ and Auger $\mathcal{A}^{(q)}$ yields accounting for fine structure for  a pulse duration of 5 fs. We find that these yields are very similar to the yields obtained in the previous section where fine structure was neglected. Thus our conclusions in the previous section regarding optimal laser parameters for observing the Auger electron spectra  up to $\mathrm{Ar^{3+}}$ still hold. Also in \fig{fig:FS_Auger} we plot the Auger yields $\mathcal{A}_{i\rightarrow j}^{(2)}$ and $\mathcal{A}_{\i \rightarrow j}^{(3)}$ for all possible $\mathrm{i}$, $\mathrm{j}$  fine structure states.

\subsection{ Auger spectra including fine structure} 
\subsubsection{One-electron Auger spectra}
\begin{figure}
\centerline{\includegraphics[width=0.98\linewidth]{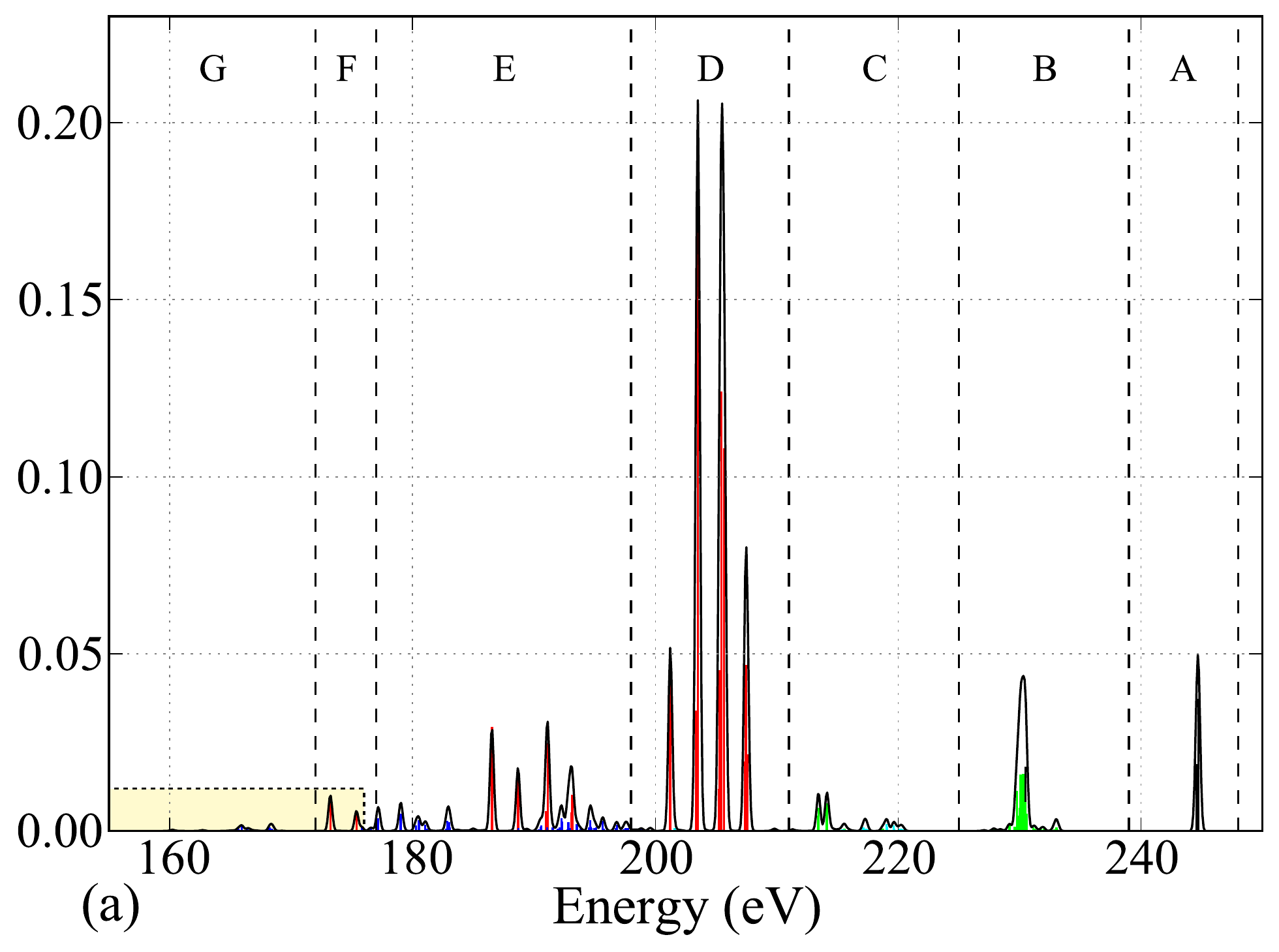}}
\centerline{\includegraphics[width=0.98\linewidth]{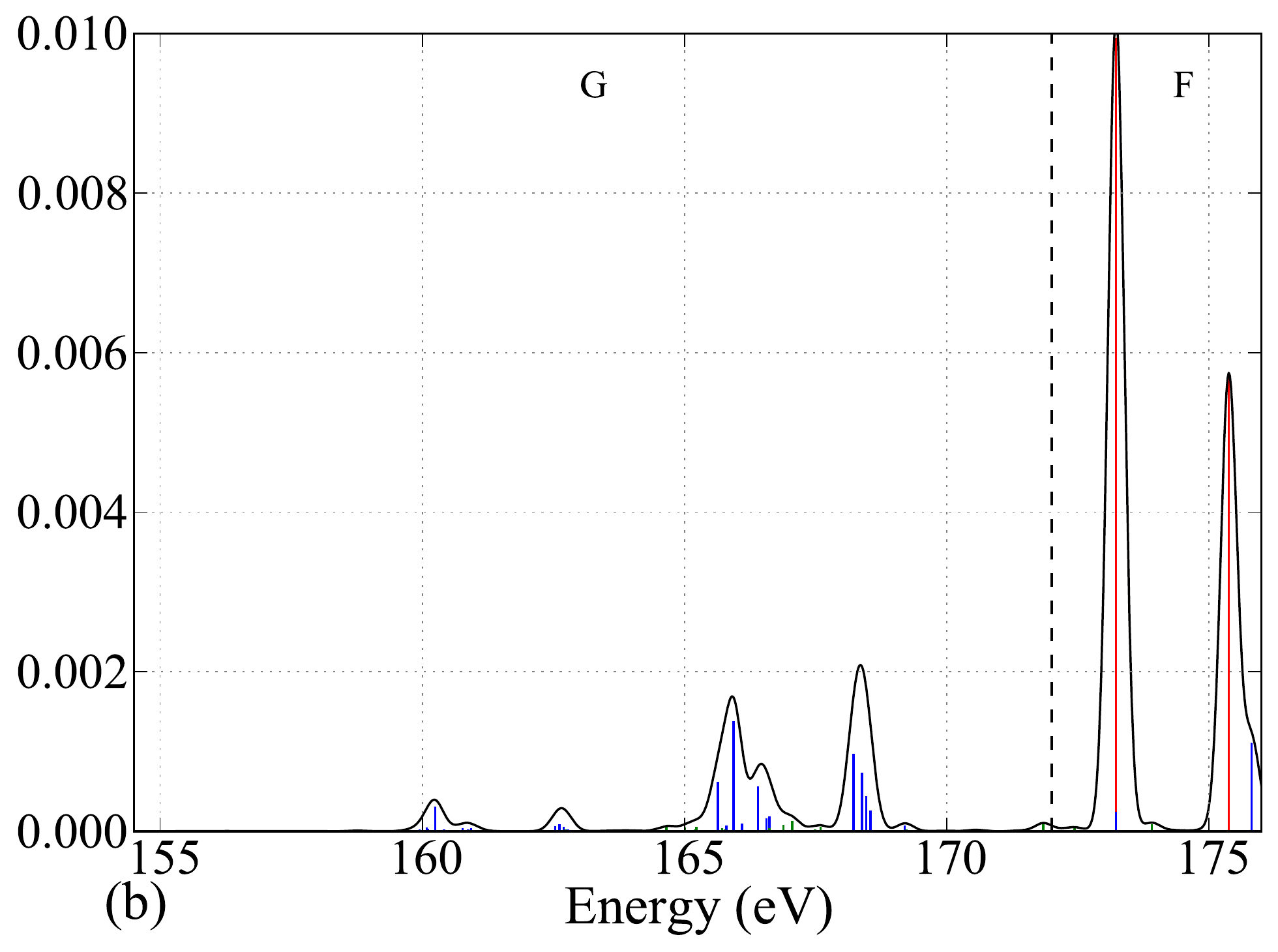}}
\caption{\label{fig:FS_fullspectrum} (Color online) The electron spectra for a 5 fs, 260 eV pulse with an intensity of $5\times10^{15}$ W cm$^{-2}$ for energies between 150 and 250 eV (a) and energies between 155 and 177 eV (b). For clarity the plot range of (b) is highlighted in yellow in (a). The peaks are convoluted by 0.37 eV FWHM Gaussian functions. 
The peaks of the photo-ionization electrons emitted during transitions from the initial states Ar (black) and Ar$^+$ (green) are in the energy ranges denoted by A, B, and C (see Table IV). 
The peaks of the Auger electrons emitted during the transitions  $\mathrm{Ar^+(2p^{-1})\rightarrow Ar^{2+}}$ (red) are in the energy ranges denoted by D, E, and F, and the ones  emitted during the transitions $\mathrm{Ar^{2+}(2p^{-1}v^{-1})\rightarrow Ar^{3+}}$ (blue) are in the energy ranges denoted by E, F, and G (see Table IV). }
\end{figure}

 In \fig{fig:FS_fullspectrum} we compute the electron spectra for a 260 eV FEL pulse with  $5\times10^{15}$ W cm$^{-2}$ intensity and  5 fs duration. Both the Auger $\mathcal{A}^{(q)}_{i\rightarrow j}$ and photo-ionization 
 $\mathcal{P}_{i\rightarrow j}^{(q)}$ yields for charges up to $q=4$ contribute to the peaks in these electron spectra. To account for the energy uncertainty  of a 5 fs pulse, which is 0.37 eV, we have convoluted the peaks  in \fig{fig:FS_fullspectrum} with Gaussian functions of 0.37 eV FWHM.
 We find that the energies of the photo-ionization electrons ejected in the transition $\mathrm{Ar}^{+}\rightarrow \mathrm{Ar}^{2+}$ (peak height $\mathcal{P}_{i\rightarrow j}^{(2)}$) are well separated from the energies of the Auger electrons ejected in the transitions $\mathrm{Ar}^{+}\rightarrow \mathrm{Ar}^{2+}$ (peak height $\mathcal{A}^{(2)}_{i\rightarrow j}$) and $\mathrm{Ar}^{2+}\rightarrow \mathrm{Ar}^{3+}$ (peak height $\mathcal{A}^{(3)}_{i\rightarrow j}$); the photo-ionization peaks are above 210 eV while the Auger peaks are below 210 eV.  
In  \fig{fig:FS_fullspectrum} and Table \ref{tab:peaks}, the energy range of the photo-ionization electrons is denoted by  A, B, C; the energy range of the Auger electrons emitted during transitions from the initial states Ar$^{+}$ and Ar$^{2+}$  are denoted by D, E, F, and E, F, and G, 
respectively. In \fig{fig:FS_fullspectrum}  we see that the Auger yields  $\mathcal{A}^{(2)}_{i\rightarrow j}$ (D, E, F) are much larger than  all other Auger yields in the same energy range. They  can thus be discerned and measured for the laser parameters under consideration. The Auger yields $\mathcal{A}^{(3)}_{i\rightarrow j}$  (E, F, G) are smaller but still visible, while the  Auger yields $\mathcal{A}^{(4)}_{i\rightarrow j}$ are too small to be discerned  in \fig{fig:FS_fullspectrum}. However, except for the energy region below 170 eV, the Auger electron spectra resulting from the transitions $\mathrm{Ar}^{2+}\rightarrow \mathrm{Ar}^{3+}$  overlap with the  Auger electron spectra resulting from the transitions $\mathrm{Ar}^{+}\rightarrow \mathrm{Ar}^{2+}$. Thus, in order to discern and be able to experimentally observe the latter Auger electron spectra  we need to consider spectra of  two electrons  in  coincidence.  We do so in what follows.
  
\begin{table}
\caption{Labeling  of energy regions in the electron spectrum shown in \fig{fig:FS_fullspectrum}. $e^-_{\rm p}$ and $e^-_{\rm A}$ stand for photo-ionization and Auger electrons, respectively. $u^{-1}$ represents a hole in any of the $2p$, $3s$ or $3p$ orbitals.
\label{tab:peaks}
}
\begin{center}
\begin{tabular}{c p{0.75\linewidth} }
\hline
Region & Transitions \\ \hline \hline
$A$ & Ar$+\hbar\omega\rightarrow$Ar$^+(3p^{-1}) +e^-_{\rm P}$  \\
$B$ & Ar$+\hbar\omega\rightarrow$Ar$^+(3s^{-1}) +e^-_{\rm P}$ \\
$B$ & Ar$^+(u^{-1}) +\hbar\omega\rightarrow$Ar$^{2+}(u^{-1}3p^{-1}) +e^-_{\rm P}$\\  
$C$ & Ar$^+(u^{-1}) +\hbar\omega\rightarrow$Ar$^{2+}(u^{-1}3s^{-1}) +e^-_{\rm P}$\\   
$D$ & Ar$^+(2p^{-1}) \rightarrow$Ar$^{2+}(3p^{-2}) + e^-_{\rm A}$\\ 
$E$ & Ar$^+(2p^{-1}) \rightarrow$Ar$^{2+}(3s^{-1}3p^{-1}) + e^-_{\rm A}$ \\ 
$F$ & Ar$^+(2p^{-1}) \rightarrow$Ar$^{2+}(3s^{-2}) + e^-_A$ \\
$E,F,G$ & Ar$^{2+}(2p^{-1}v^{-1}) \rightarrow$Ar$^{3+}(v^{-3})+ e^-_A$  \\
 \hline
\end{tabular}
\end{center}
\end{table}%

\subsubsection{Two-electron coincidence Auger spectra}
We now consider the electron spectra resulting from the transitions:
\begin{equation}
\mathrm{Ar} + \hbar\omega\rightarrow \mathrm{Ar}^{+}(2p^{-1})+e^-_{P}  \rightarrow \mathrm{Ar}^{3+} + e^-_{P} + e^-_{B} + e^-_{C}
\label{transition}
\end{equation}
The photo-ionization electron $\mathrm{e_{P}^{-}}$  has an energy of 12.3 eV for $\mathrm{Ar}^{+}(2p^{-1}_{3/2})$ and 10.2 eV for  $\mathrm{Ar}^{+}(2p^{-1}_{1/2})$. This energy  is very different from the energies of electrons $\mathrm{e_{B}^{-}}$ and $\mathrm{e_{C}^{-}}$. It thus suffices to plot in coincidence  the energies of  electrons $\mathrm{e_{B}^{-}}$ and $\mathrm{e_{C}^{-}}$. We note that  many coincidence experiments have been performed with synchrotron radiation \cite{Alkemper:1997,Lablanquie:2007,Lablanquie:multi:2011,Huttula:2013}. 
While some coincidence experiments have been performed with FEL radiation \cite{Kurka:2009,Rudenko:2010} the low repetition rate poses a challenge. Advances in FEL sources should overcome such challenges in the near future.

\begin{figure}
\centerline{\includegraphics[width=1.\linewidth]{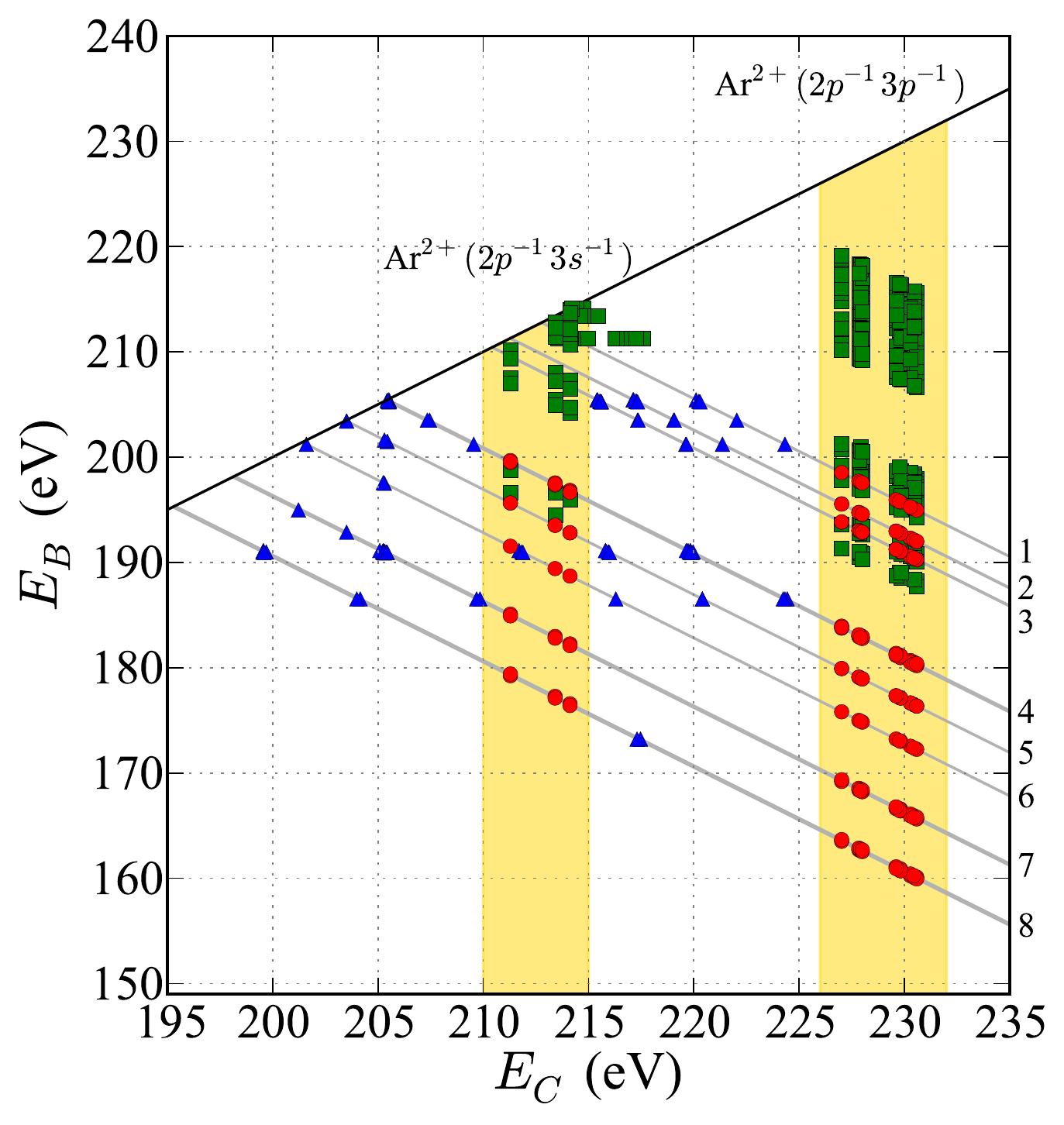}}
\caption{\label{fig:FS_coincidence1} (Color online) Two-electron coincidence spectra for Ar$^+(2p_{3/2}^{-1})\rightarrow$Ar$^{3+}(v^{-3})$ generated by a 5 fs, $5\times10^{15}$ W cm$^{-2}$ FEL pulse. We show the spectrum below the $E_B=E_C$ line for the PPP (green squares), PAP (blue triangles) and PPA (red circles) transition sequences, see text for details. The Ar$^{3+}(3p^{-3})$ final fine structure states are labeled as  1:$^4S$, 2:$^2D$, 3:$^2P$, the Ar$^{3+}(3s^{-1}3p^{-2})$  states as 4:$^4P$, 5:$^2D$, 6:$^2S$, 7:$^2P$, and the Ar$^{3+}(3s^{-2}3p^{-1})$ state as 8:$^2P$. The energy range of the PPA transition sequences   $\mathrm{Ar^{+}(2p^{-1}) \rightarrow Ar^{2+}(2p^{-1}3s^{-1}) \rightarrow Ar^{3+}}$ and  $\mathrm{Ar^{+}(2p^{-1}) \rightarrow Ar^{2+}(2p^{-1}3p^{-1}) \rightarrow Ar^{3+}}$ are highlighted by yellow.
}
\end{figure}

\begin{figure}
\centerline{\includegraphics[width=1.\linewidth]{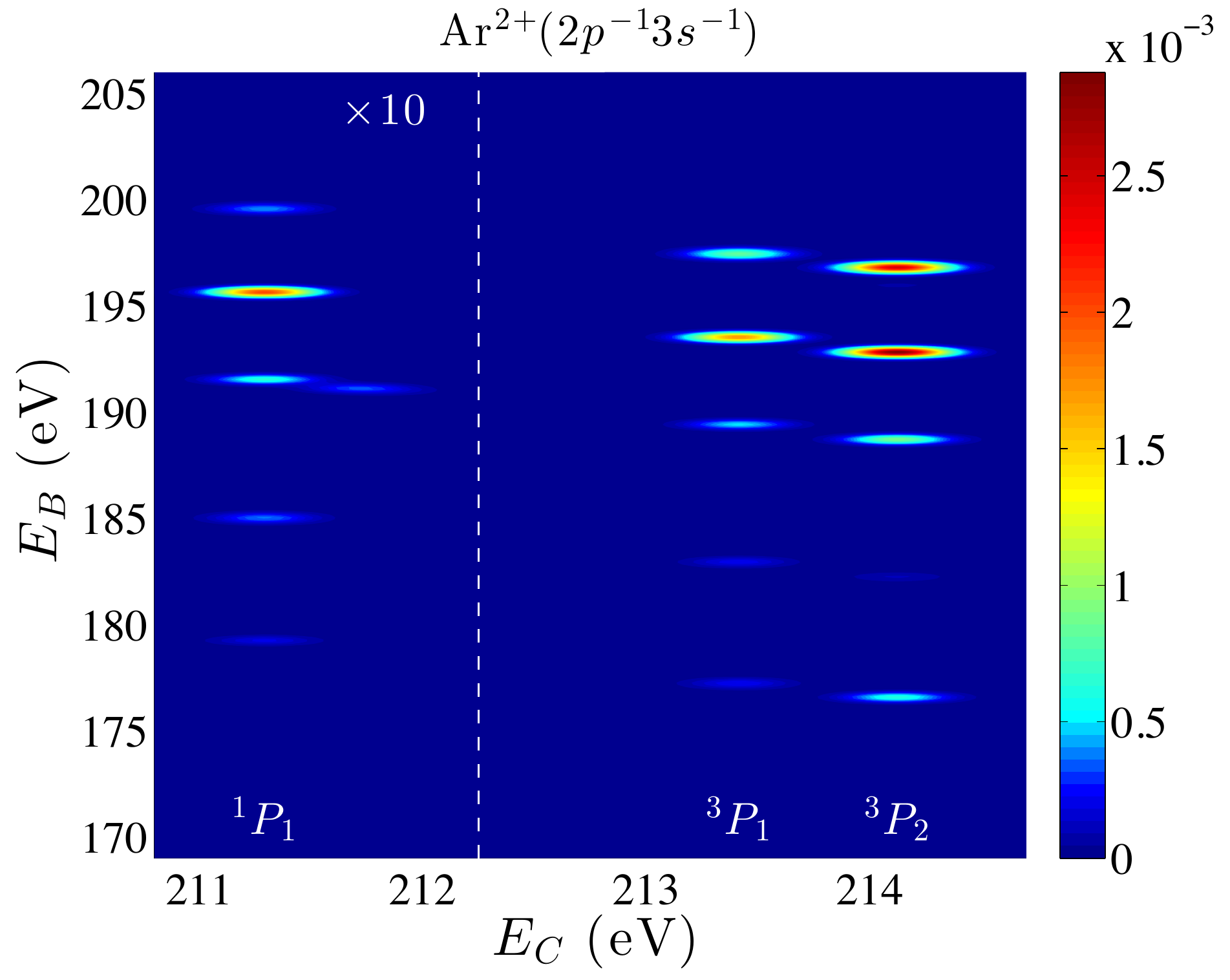}}
\caption{\label{fig:FS_coincidence2} 
(Color online) Two-electron coincidence spectra for  $\mathrm{Ar^{+}(2p^{-1}) \rightarrow Ar^{2+}(2p^{-1}3s^{-1}) \rightarrow Ar^{3+}}$ for the $\mathrm{Ar^{2+}(2p^{-1}3s^{-1})}$ fine structure states  $\mathrm{^1P_1}$,  $\mathrm{^3P_1}$ and  $\mathrm{^3P_2}$.
The  peaks of the spectra corresponding to the Ar$^{2+}(2p^{-1}3s^{-1};^1P_1)$ fine structure state, to the left of the vertical dashed-white line, are much smaller than the rest of the spectra and we have  thus multiplied them by a factor of 10 so that they are visible.
The coincidence peaks have been convoluted by 0.37 eV FWHM Gaussian functions.}
\end{figure}

In  \fig{fig:FS_coincidence1} we plot in coincidence the  energies of electrons $\mathrm{e^-_{B}}$ and $\mathrm{e^-_{C}}$.  
Specifically, \fig{fig:FS_coincidence1}  corresponds to the  $\mathrm{ Ar^{+}(2p^{-1}_{3/2})}$ fine structure state in \eq{transition}. We only show the spectrum that lies below the line $E_B=E_{C}$ 
(black solid line), with $E_{B}$ the energy of electron  $\mathrm{e_{B}^{-}}$ and $E_C$ the energy of electron  $\mathrm{e_{C}^{-}}$. Since the two electrons are indistinguishable, the remaining spectrum can be obtained  by a reflection with respect to the line $E_B=E_{C}$ of the  spectrum  shown in \fig{fig:FS_coincidence1}. 
 From \eq{transition} it follows that each line with $E_{B}+E_{C}=constant$, grey lines in   \fig{fig:FS_coincidence1}, scans the spectra of electrons emitted from transitions in \eq{transition} through any possible fine structure state of Ar$^{2+}$ to the same  fine structure state of Ar$^{3+}$.  The spectra of the electrons emitted from the transitions in \eq{transition} can be labelled according to the sequence of photo-ionization (P) and Auger processes (A) involved while transitioning from Ar to Ar$^{3+}$: 
 PPA (red in \fig{fig:FS_coincidence1}), PPP (green) and PAP (blue). Our goal is to retrieve the Auger electron spectra corresponding to the transitions $\mathrm{ Ar^{2+} \rightarrow Ar^{3+}}$. These latter spectra are the ones labelled as PPA in  \fig{fig:FS_coincidence1}; we highlight the energy range of the $\mathrm{e_{B}^{-}}$ and $\mathrm{e_{C}^{-}}$  electrons emitted in the PPA  transition sequences $\mathrm{Ar^{+}(2p^{-1}) \rightarrow Ar^{2+}(2p^{-1}3s^{-1}) \rightarrow Ar^{3+}}$ and  $\mathrm{Ar^{+}(2p^{-1}) \rightarrow Ar^{2+}(2p^{-1}3p^{-1}) \rightarrow Ar^{3+}}$. Thus to be able to retrieve the Auger electron spectra associated with the transitions $\mathrm{ Ar^{2+} \rightarrow Ar^{3+}}$ we must be able to discern the PPA from the PPP and the PAP transition sequences. We see
  that in the highlighted area in \fig{fig:FS_coincidence1} there is some small overlap of the PPA with the PPP and PAP  sequences. However, we find that the height of the peaks of the PPA transition sequences are much larger than the height of the peaks of the PPP and PAP transition sequences. Specifically,  the total Auger yield $\mathcal{A}^{(3)}$ associated with the PPA transition sequences is roughly 5 times larger than the 
photo-ionization yield $\mathcal{P}^{(3)}_{PAP}$ corresponding to the PAP transition sequences   and 10 times larger than the photo-ionization yield $\mathcal{P}^{(3)}_{PPP}$ corresponding to the PPP transition sequences, with $\mathcal{P}^{(3)}_{PAP}+\mathcal{P}^{(3)}_{PPP}=\mathcal{P}^{(3)}$. To show that this is indeed the case we show in \fig{fig:FS_coincidence2} the contour plot of  the two-electron coincidence spectra associated with the highlighted area in \fig{fig:FS_coincidence1} corresponding to the transitions $\mathrm{Ar^{+}(2p^{-1}) \rightarrow Ar^{2+}(2p^{-1}3s^{-1}) \rightarrow Ar^{3+}}$. Note that the height of the peaks in \fig{fig:FS_coincidence2} is given by $\mathcal{A}^{(3)}_{j\rightarrow k}$ or  $\mathcal{A}^{(3)}_{j(i)\rightarrow k}$  (see discussion in section IIA) for the PPA transition sequences while the height is   $\mathcal{P}^{(3)}_{j\rightarrow k}$ or  $\mathcal{P}^{(3)}_{j(i)\rightarrow k}$ for the PPP and PAP transition sequences. Each coincidence peak has been convoluted by a 0.37 eV FWHM Gaussian function.  We find that all except one of the observable peaks in \fig{fig:FS_coincidence2} are due to PPA transition sequences; the small height peak at $(E_C=211.7,E_B=190)$ is due to a PAP sequence. We have thus demonstrated that we can retrieve from the two-electron coincidence spectra the Auger electron spectra associated with the  transitions $\mathrm{ Ar^{2+} \rightarrow Ar^{3+}}$.  
We note that a similar discussion and conclusions  hold for the Auger spectra corresponding to the  $\mathrm{ Ar^{+}(2p^{-1}_{1/2})}$ fine structure state in \eq{transition}.

Finally we note that our calculations neglect satellite structure. That is, we do not account for Auger transitions where one electron fills in the 2p hole, another one escapes to the continuum while a third one is promoted to an excited state. The main (larger) satellite Auger yields we are neglecting  are most likely due to the  Ar$^+(2p^{-1})\rightarrow$Ar$^{2+}(3s^{-1}3p^{-1})$ transition\cite{Pulkkinen:1996}. However, these satellite yields are  smaller than the main Auger yields for this transition. In addition, these satellite Auger yields would only contribute to the part of the spectrum corresponding to PAP transition sequences in the energy region  $E_B=$170 -180 eV and $E_C=$210-220 eV. But as  we discussed above the contribution to the electron spectra from PAP transition sequences   is  smaller than the contribution from the PPA transition sequences. Thus our approximation is justified.

\section{Conclusions}

We have explored the interplay of photo-ionization and Auger transitions in Ar when interacting with a 260 eV  FEL pulse. Solving the rate equations we have explored the dependence of the ion and Auger yields on the laser parameters accounting, at first, only for the electron configuration of the ion states. We have found that an FEL pulse of roughly 5 fs  duration and $5\times 10^{15}$ Wcm$^{-2}$ intensity is optimal for retrieving Auger electron spectra up to Ar$^{3+}$. 
Secondly, we have account for the fine structure of the ionic states and have truncated the rate equations to include states only up to $\mathrm{Ar^{4+}}$.
We have  shown how the Auger electron spectra of   $\mathrm{Ar^{+}\rightarrow Ar^{2+}}$ can be retrieved.  We have also shown that the Auger electron spectra of  $\mathrm{Ar^{2+}\rightarrow Ar^{3+}}$ can also be retrieved when two electrons are considered in coincidence.  We have thus demonstrated that interaction with FEL radiation is a possible route for retrieving Auger electron spectra. We believe that our work will stimulate further theoretical and experimental studies along these lines.

%
%

\section{Acknowledgments}

The authors are grateful to Prof. P. Lambropoulos for initial motivation and valuable discussions.
A.E. acknowledges support from EPSRC under Grant No. H0031771 and J0171831  and use of the Legion computational resources at UCL.

\bibliography{AOGW_all.bib}

\end{document}